\documentclass[conference]{IEEEtran}
\IEEEoverridecommandlockouts
\usepackage{cite}
\usepackage{amsmath,amssymb,amsfonts}
\usepackage{algorithmic}
\usepackage{graphicx}
\usepackage{textcomp}
\usepackage{xcolor}
\usepackage{CJKutf8}
\usepackage{color, xcolor}
\usepackage{cite}
\usepackage{ulem}
\usepackage{diagbox}
\usepackage{gbt7714}

\def\BibTeX{{\rm B\kern-.05em{\sc i\kern-.025em b}\kern-.08em
		T\kern-.1667em\lower.7ex\hbox{E}\kern-.125emX}}

\makeatletter
\newcommand{\linebreakand}{%
  \end{@IEEEauthorhalign}
  \hfill\mbox{}\par
  \mbox{}\hfill\begin{@IEEEauthorhalign}
}
\makeatother

\begin{document}
	
	\title{Dual-modality Smart Shoes for Quantitative Assessment of Hemiplegic Patients' Lower Limbs' Muscle Strength \\
% 		\thanks{Identify applicable funding agency here. If none, delete this.}
	}
	
	\author{
	    
 		\IEEEauthorblockN{1\textsuperscript{st} Huajun Long}
 		\IEEEauthorblockA{\textit{Dept. of Data Science} \\
 		    \textit{University of Science and}\\
 			\textit{Technology of China(USTC)}\\
 			Hefei, China \\
 			sa20229062@mail.ustc.edu.cn}
 		\and
 		\IEEEauthorblockN{2\textsuperscript{nd} Jie Li}
 		\IEEEauthorblockA{
 		    \textit{Department of Neurology}\\
 			\textit{The First Affiliated Hospital of USTC}\\
 			Hefei, China \\
 			lijie01@mail.ustc.edu.cn}
 		\and
 		\IEEEauthorblockN{3\textsuperscript{rd} Rui Li}
 		\IEEEauthorblockA{\textit{Department of Neurology} \\
 			\textit{The First Affiliated Hospital of USTC}\\
 			Hefei, China \\
 			348102720@qq.com}
			
         \linebreakand
 		\IEEEauthorblockN{4\textsuperscript{th} Xinfeng Liu}
 		\IEEEauthorblockA{\textit{Department of Neurology} \\
 			\textit{The First Affiliated Hospital of USTC}\\
 			Hefei, China \\
 			xfliu2@ustc.edu.cn}
 		\and
 		\IEEEauthorblockN{5\textsuperscript{th} Jingyuan Cheng}
 		\IEEEauthorblockA{\textit{Dept. of Computer Science and Technology} \\
 			\textit{University of Science and Technology of China}\\
 			Hefei, China \\
 			jingyuan@ustc.edu.cn}
	}

	\maketitle
	
	\begin{abstract}
		
		Stroke can lead to the impaired motor ability of the patient's lower limbs and hemiplegia.  Accurate assessment of the lower limbs' motor ability is important for diagnosis and rehabilitation. To digitalize such assessment so that each test can be traced back any time and subjectivity can be avoided,  we test how dual-modality smart shoes equipped with pressure-sensitive insoles and inertial measurement units can be used for this purpose. A 5m walking test protocol, including the left and right turns, is designed. Data are collected from 23 patients and 17 healthy subjects. For the lower limbs' motor ability, the tests are observed by two physicians and assessed using the five graded Medical Research Council scale for muscle examination. The average of two physicians' scores for the same patient is used as the ground truth. Using the feature set we developed, 100\% accuracy is achieved in classifying the patients and healthy subjects. For patients' muscle strength, a mean absolute error of 0.143 and a maximum error of 0.395 is achieved using our feature set and the regression method, closer to the ground truth than the scores from each physician~(mean absolute error: 0.217, maximum error: 0.5). We thus validate the possibility of using such smart shoes to objectively and accurately evaluate the lower limbs' muscle strength of the stroke patients.
		
	\end{abstract}
	
	
	\section{Introduction}
	\label{sec:intruduction}
	
	With approximately 12.2 million new cases worldwide in 2019, stroke has become one of the most prevalent diseases~\cite{feigin2021global}. It is difficult to cure, prone to recurrence and slow to recover. The patients need regular clinical assessments to measure the rehabilitation progress after the acute phase of in-hospital treatment~\cite{centers2018national}. One of the main purposes of rehabilitation is to improve independent mobility, where the assessment of lower limbs' muscle strength is essential~\cite{andrews2003short}.

Two types of methods are practised for the muscle strength assessment: the medical scales or the equipment. The MRC scale~(Medical Research Council scale for muscle examination)~\cite{gregson2000reliability} is often used to grade the patients' lower limbs' muscle strength. The patient is observed by a physician and given a score~(from 0 to 5). The results might be biased, as different physicians might give different scores to the same patient. To digitalize the assessment, equipments can be used, such as muscle testing dynamometer, optical motion capture system or force plate~(more details in section~\ref{subsec:device}). The whole process can be traced back, and subjectivity can be avoided. 

We take the double-modality smart shoes as our measuring device, as they are small in size, easy to use, and comparatively inexpensive, thus own the potential for continuous assessment at home. To map the large amount of test data to the lower limb's motor ability, we follow the general data mining process, building first the general feature set, then converting the feature set into the motor ability using the regression method. To obtain the ground truth, two physicians observed the tests and provided their independent scores using the MRC scale.
	
Our work validates the possibility of using dual-modality smart shoes to objectively  evaluate the stroke patients' lower limbs' muscle strength. The main contributions are:
\begin{itemize}
	\item We extended the 5m walk test and included the left and right turns, and demonstrate with the muscle strength's regression results that this extension is essential.
	\item  We proposed the dual-modality fusion features and proved the importance of these newly proposed features.
	\item We proposed a two-step evaluation method. A subject is first classified as patient or healthy. If classified as a patient, the lower limbs' muscle strength is then calculated using our feature set and the regression method. Based on the data collected from 23 patients and 17 healthy subjects, 100\% classification accuracy is achieved. The regression result~(mean absolute value error: 0.143, maximum error: 0.395) is even closer to the ground truth than the scores of individual physicals~(mean absolute value error: 0.217, maximum error: 0.500).
	
\end{itemize}

\section{Related Work}
	\label{sec:related_work}
	
\subsection{Devices for assessing lower body mobility}
	\label{subsec:device}
	
Various devices have been developed for assessing motor abilities: muscle strength meter, optical motion capture system, force plate, inertial sensor, and smart shoes. Mentiplay et al.~\cite{mentiplay2015assessment} demonstrated that handheld plyometrics can measure muscle strength reliably and effectively. However, the measurement requires the assistance of other people, and only  muscle strength at rest can be obtained. Rastegarpanah et al.~\cite{rastegarpanah2018targeting} used the VICON MX force plate system with an optical motion capture system to analyze spatial-temporal gait parameters in healthy and stroke groups.  Wang et al.~\cite{wang2021effect} used F-scan insoles to study temporal changes in each gait phase over four weeks in hemiplegic patients. Although the combination of an optoelectronic system and a pressure carpet are commonly used, the devices are expensive and require the professional operation, making long-term continuous monitoring difficult. 

Wearable systems enable measurement at any place. For example,Yang et al.~\cite{yang2013estimation} attached multiple Inertial Measurement Units~(IMUs) to the legs to study walking speed on the hemiplegic side versus the non-hemiplegic side. Smart shoes are another option, as people normally walk with shoes on. Smart shoes equipped with IMU and/or pressure sensors can be cheap, easy to use, and suitable for monitoring individual gait patterns for a long time. In Table~\ref{table:smart_shoes}, we provide an overview of the existing smart shoes and the shoe we've developed and used in our research. Our system features 6-axial IMUs and a pressure-sensitive matrix covering the whole plantar, providing multi-modality, detailed data for gait monitoring.
 
 	\begin{table}
 	\setlength\tabcolsep{3pt}
		\centering
		\caption{Smart shoe systems}
		\begin{tabular}{|c|c|c|c|c|}
			\hline
			Smart insole & IMU & Sensor No. & Battery life & Sampling rate \\ \hline
			Moticon Insole~\cite{ReGo} & Yes & 16 & N/A & 100Hz \\ \hline
			Fazio et al.~\cite{de2021development} & Yes & 48 & N/A & 100Hz \\ \hline   
			F-Scan~\cite{F-Scan} & No & 960 & 2h  & 100Hz \\ \hline
			Pedar-X Insole~\cite{lin2016smart} & No & 99 & 4.5h  & 100Hz \\ \hline
			Digitsole~\cite{digitsole} & No & N/A & 7h or 8h  & 208Hz \\ \hline
			Ours & Yes & 400 & 12h  & 60Hz \\ \hline
		\end{tabular}
		\label{table:smart_shoes}
	\end{table}
 	
\subsection{Experimental design}
	\label{experimental_design}
Experiment design plays an important role in data gathering. Current test protocols include treadmill walking test~\cite{chen2005gait}, up and down stairs test~\cite{laudanski2013measurement}, long straight corridor walking test~\cite{yang2013estimation,hodt2008should}, and times up and go test~\cite{bonnyaud2015spatiotemporal}. Walking on a treadmill allows strict control of the walking speed but requires an extra device. Walking up and down stairs test can better test the walking ability under extreme conditions, but also brings greater safety risks. The 5m walk test is close to daily life, easy to operate, and is of less safety risk. Because Bonnyaud et al.~\cite{bonnyaud2015spatiotemporal} found that walking parameters during patient turning were related to lower limbs' motor ability, we extend the 5m walk test and include the right and left turnings.
	
\subsection{Data processing algorithms}
	\label{subsec:algorithm_and_results}
	
We take the early works using pressure sensors and IMU as the base of our algorithms. Galli et al.~\cite{galli2010gait} calculated the walking speed of both feet and the difference in acceleration and angular velocity at the joints to quantitatively compare the hemiplegic children's walking ability. Wang et al.~\cite{wang2020imu} calculated the acceleration parameters from the inertial sensors and proposed the gait normal index to evaluate the lower limbs' motor ability. Echigoya et al.~\cite{echigoya2021changes} calculated the changes in the center of plantar pressure and analyzed the important parameters that allow independent walking. Chisholm et al.~\cite{chisholm2011inter} calculated the spatial-temporal parameters of the center of plantar pressure and analyzed the correlation with the severity of the sensorimotor impairment by comparing the hemiplegic side of the patient with the non-hemiplegic side. Chen et al.~\cite{chen2005gait} calculated gait phase duration, walking speed, step length and step width. We borrowed some of these parameters. Because our hardware offers two sensing modalities, we are able to propose the dual-modality fusion features, whose high importance will be discussed further in section~\ref{subsec:feature}.
	
\section{Experimental Setup and Acquisitions}
	\label{sec:experimental_setup}

\begin{figure}[htbp]
		\centerline{\includegraphics[width=0.5\textwidth]{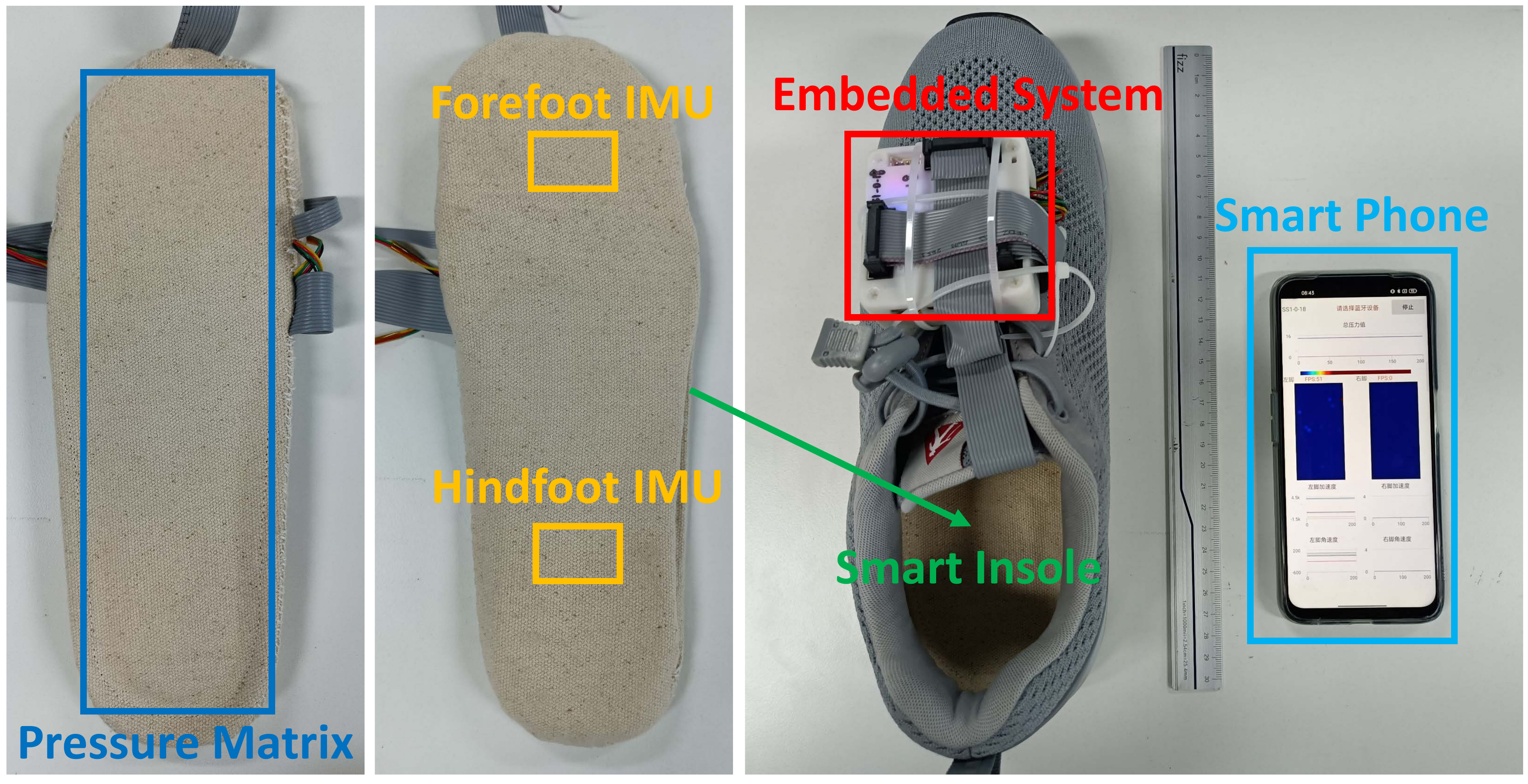}}
		\caption{Our smart shoe prototype, plus a mobile phone for data storage}
		\label{fig:devices}
	\end{figure}

Our smart shoe system is shown in Fig.~\ref{fig:devices}. Each shoe is equipped with a textile pressure sensing insole, covering the whole planta~(horizontal $16 \times$ vertical $25$) and two 6-axial IMUs~(tri-axial accelerometer plus tri-axial gyroscope) located at the forefoot and hindfoot positions. The sample rate is 60$Hz$ and data are transmitted to the mobile phone via Bluetooth.
	
% Data acquisition was conducted in The First Affiliated Hospital of University of Science and Technology of China. The test subjects were instructed to walk at a self-selected speed after putting on both smart shoes, following three routines:

Data acquisition was conducted in xxxxxx. The test subjects were instructed to walk at a self-selected speed after putting on both smart shoes, following three routines:
	\begin{itemize}
		\item \textit{Straight}: Go straight from the starting point for 5m, turn around, walk straight back to the starting point, stop.
		\item \textit{Right-turning}: From the starting point along the edge of a square~(all the sides are of 5m length), turn right at the corners, stop at the starting point.
		\item \textit{Left-turning}: Similar to \textit{right-turning}, only the turning direction is left at the corners.
	\end{itemize}

In total, 23 stroke patients and 17 healthy subjects were involved in this study~(details in table~\ref{tab:participants}). The stroke patients have hemiplegia on one side and no significant symptoms on the other side, are conscious and able to walk at least with assistance. The healthy subjects have no stroke symptoms, normal lower limbs' motor ability and clear consciousness. For each patient, two clinically experienced physicians used the MRC scale to give individual evaluation scores on the lower limbs' muscle strength. The scores ranged from 3- to 5-. We map 3-, 3, 3+, 4-, 4, 4+, 5- to 2.67, 3.00, 3.33, 3.67, 4.00, 4.33, 4.67, and use the average scores of the two physicians as the ground truth. 
	
	\begin{table}[!ht]
		\caption{The clinical characteristics of the participants}
		\centering
		\begin{tabular}{|p{60pt}|p{60pt}|p{60pt}|}
			\hline
			 & Stroke patients & Healthy subjects\\ \hline
			Male/Female & 14/9 & 10/7  \\ 
			Age~(years) & 62 $\pm$ 11 & 60 $\pm$ 9  \\ 
			Paresis side~(L/R) & 5/18 & N/A  \\  \hline
		\end{tabular}
		\label{tab:participants}
	\end{table}

\section{Data Processing and Feature Extraction}
	\label{sec:data_processing_and_feature_extraction}
		
\begin{figure*}[htbp]
		\centerline{\includegraphics[width=0.9\textwidth]{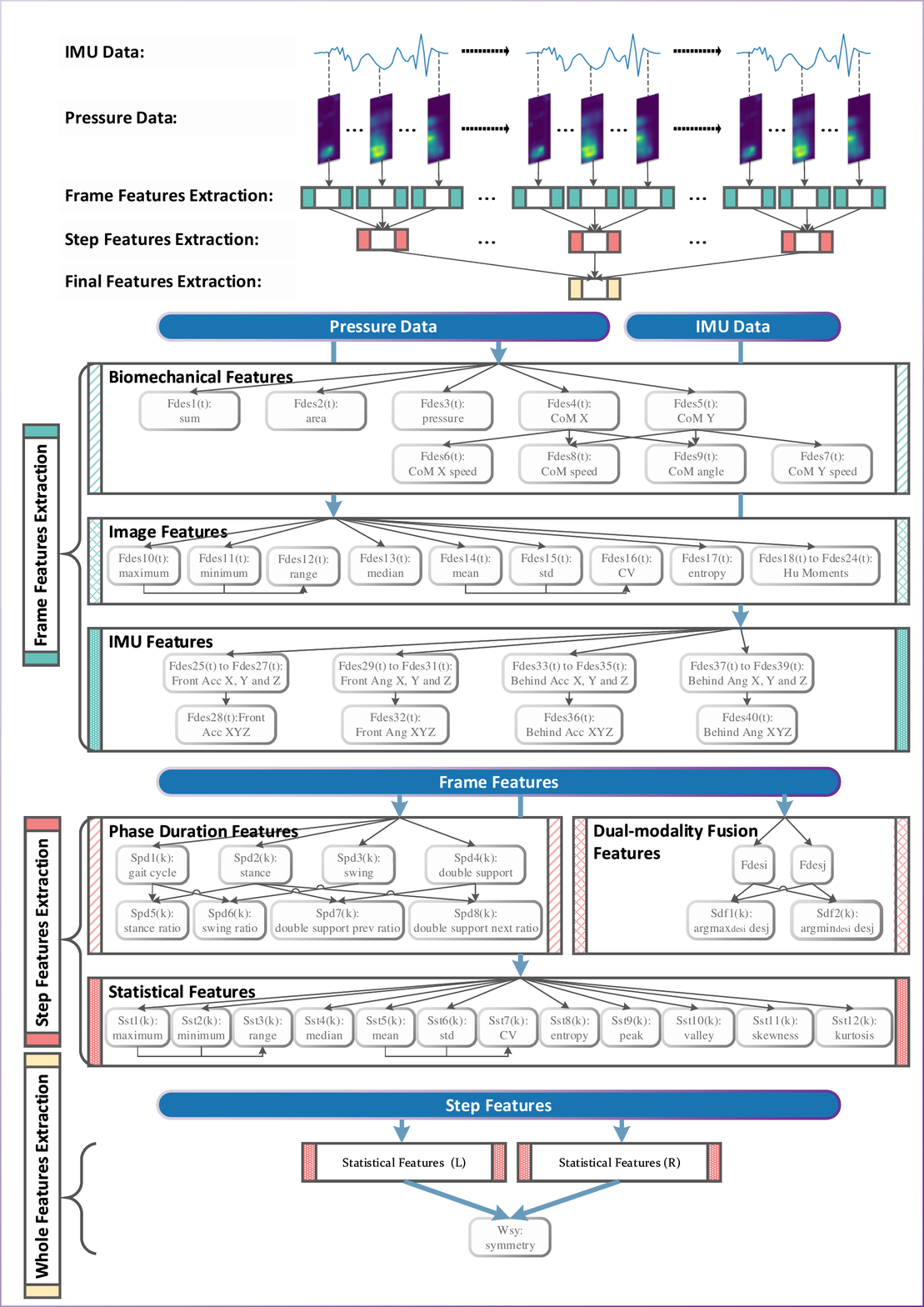}}
		\caption{The feature extraction workflow}
		\label{fig:feature_extraction_process}
	\end{figure*}

	We follow the classic data mining process. The data from the pressure insole and IMU of each shoe are first pre-processed, and then the data from both shoes are synchronized and fed into the general feature extraction flow~(shown in Fig.~\ref{fig:feature_extraction_process}) to obtain frame features, step features and whole features. These features are used later as input for classification and regression algorithms, described in section~\ref{sec:results}.
	
\subsection{Pre-processing}
	\label{subsec:preprocessing}
	
	The pressure data go through linear interpolation, up-sampling and Gaussian smoothing, while the IMU data are only linear-interpolated. 
	
	\subsubsection{Linear interpolation}
	
	Data could get lost in the wireless transmission. To ensure a constant sample rate, data are repaired using linear interpolation. Given $n$ samples between time $t_x$ and $t_y$ are lost, from the data we get before and after this period $data_{tx}$, $data_{ty}$, the $i^{th}$ lost data are recovered as:
	
	\begin{equation}
		data_{ti} = \frac{i}{n + 1} (data_{ty} - data_{tx}) + data_{tx}
	\end{equation}
	
	\subsubsection{Up-sampling}
	Up-sampled pressure image generates more accurate image features~\cite{sundholm2014smart}. The pressure images are thus up-sampled to $32 \times 100$ using bilinear interpolation.
	
	\subsubsection{Gaussian smoothing}
	To obtain a smoother distribution, A $5 \times 5$ Gaussian filter is applied on each pressure image.
	
	\subsection{Frame features extraction}
	\label{subsec:frame_feature}
	
	The pressure and IMU data at each sample is defined as a "frame", from which the frame features are extracted. 
	
	For the pressure data, Zhou et al.~\cite{zhou2019tpm} proposed the TPM feature set~(743 features). Guo et al.~\cite{guo2022lwtool} further expanded the features' number to 1830. We borrow 23 of the frame features that were confirmed to be associated with lower limbs' motor ability of stroke patients~\cite{wang2020imu, echigoya2021changes,chisholm2011inter,chen2005gait}, supplemented with $Fdes_{16}(t)$, which measures the overall variance in the pressure images. These features are grouped into the bio-mechanical features~($Fdes_1-Fdes_9$) and the image features~($Fdes_{10}-Fdes_{24}$).
	
	For the IMU data, we extracted from the raw signal 16 frame features~($Fdes_{25}-Fdes_{40}$). 
	
	The symbols used are defined in Table~\ref{table:summary_of_symbols} and will be used throughout the paper. 
	
	\begin{table}[!ht]
		\caption{The summary of the symbols in the paper}
		\centering
		\begin{tabular}{p{30pt}|p{195pt}}
			\hline
			Symbol & Description \\ \hline
			$m$ & the number of columns in the pressure image, $m=32$ \\
			$n$ & the number of rows, $n=100$ \\
			$x$ & the column index, $x \in [1,m]$ \\
			$y$ & the row index, $y \in [1,n]$ \\
			$t$ & the frame index \\
			$k$ & the step index \\
			$p(x, y, t)$ & the pressure at point $(x, y)$ in the $t^{th}$ frame\\
			$M$ & the number of pixels in the pressure image, $M=32 \times 100$ \\
			$T$ & the number of frames in one gait cycle \\ \hline
		\end{tabular}
		\label{table:summary_of_symbols}
	\end{table}
	
	\subsubsection{Biomechanical features} In total 9 features.

	\begin{itemize}
		\item Total force~($Fdes_1(t)$):
		\begin{equation}
			Fdes_1(t) = \sum\limits_{x,y}^Mp(x,y,t)
		\end{equation}
		\item Area~($Fdes_2(t)$)~(the count of pixels that are above the $\delta$):
		\begin{equation}
			\delta = 0.7 \cdot mean(p(x, y, t)) + 0.3\cdot min(p(x, y, t))
		\end{equation}
		where the mean and min operation take into account all the $p(x,y,t)$ within the whole test.
		\item Average pressure~($Fdes_3(t)$):
		\begin{equation}
			Fdes_3(t) = Fdes_1(t) / Fdes_2(t)
		\end{equation}
		\item The centre of mass~(CoM)~($Fdes_4(t)$ and $Fdes_5(t)$):
		\begin{equation}
			\begin{cases}
				Fdes_4(t) = \sum_{x=1}^m \sum_{y=1}^n x \cdot p(x, y, t) / Fdes_1(t) \\
				Fdes_5(t) = \sum_{x=1}^m \sum_{y=1}^n y \cdot p(x, y, t) / Fdes_1(t)
			\end{cases}
		\end{equation}
		\item The CoM's speed, its magnitude~($Fdes_8(t)$) and the projections on the horizontal~($Fdes_6(t)$) and vertical direction~($Fdes_7(t)$):
		\begin{equation}
			\begin{cases}
				Fdes_6(t) = Fdes_4(t)-Fdes_4(t-1)\\
				Fdes_7(t) = Fdes_5(t)-Fdes_5(t-1)\\
				Fdes_8(t) = \sqrt{Fdes_6(t)^2+Fdes_7(t)^2}
			\end{cases}
		\end{equation}
		\item The CoM's moving direction~($Fdes_9(t)$):
		\begin{equation}
			Fdes_9(t) = arctan(Fdes_6(t) \, / \, Fdes_7(t))
		\end{equation}
	\end{itemize}
	
	\subsubsection{Image features} In total 15 features.
	\begin{itemize}
		\item Maximum, minimum, range, median, mean, standard deviation of the pressure image $p(x,y,t)$~($Fdes_{10}(t)$ to $Fdes_{15}(t)$). Specially, ''range'' is defined as:
		\begin{equation}
			Fdes_{12}(t) = Fdes_{10}(t) - Fdes_{11}(t)
		\end{equation}
		\item Coefficient of variation~($Fdes_{16}(t)$):
		\begin{equation}
			Fdes_{16}(t) = Fdes_{15}(t)/Fdes_{13}(t)
		\end{equation}
		\item Information entropy~($Fdes_{17}(t)$);
		\item The image's Hu-Moments~($Fdes_{18}(t)$ to $Fdes_{24}(t)$)~\cite{hu1962visual}.
	\end{itemize}
	
	\subsubsection{IMU features} in total 16 features. 
	\begin{itemize}
		\item Acceleration of the forefoot IMU, its magnitude~($Fdes_{28}(t)$) and the projections on the x, y, z axis~(in the local coordinate)~($Fdes_{25}(t)$ to $Fdes_{27}(t)$):
		\begin{equation}
			Fdes_{28}(t) = \sqrt{Fdes_{25}(t)^2 + Fdes_{26}(t)^2 + Fdes_{27}(t)^2}
		\end{equation}
		\item Angular velocity, its magnitude~($Fdes_{32}(t)$)  and the projections~($Fdes_{29}(t)$, $Fdes_{30}(t)$ and $Fdes_{31}(t)$):
		\begin{equation}
			Fdes_{32}(t) = \sqrt{Fdes_{29}(t)^2 + Fdes_{30}(t)^2 + Fdes_{31}(t)^2}
		\end{equation}
		\item Acceleration~($Fdes_{33}(t)$-$Fdes_{36}(t)$) and angular velocity~($Fdes_{37}(t)$-$Fdes_{40}(t)$) of the hindfoot IMU, similar to that of the forefoot.		
	\end{itemize}
	
\subsection{Step features extraction}
	\label{subsec:step_features}
	
\begin{figure*}[htbp]
		\centerline{\includegraphics[width=0.75 \textwidth]{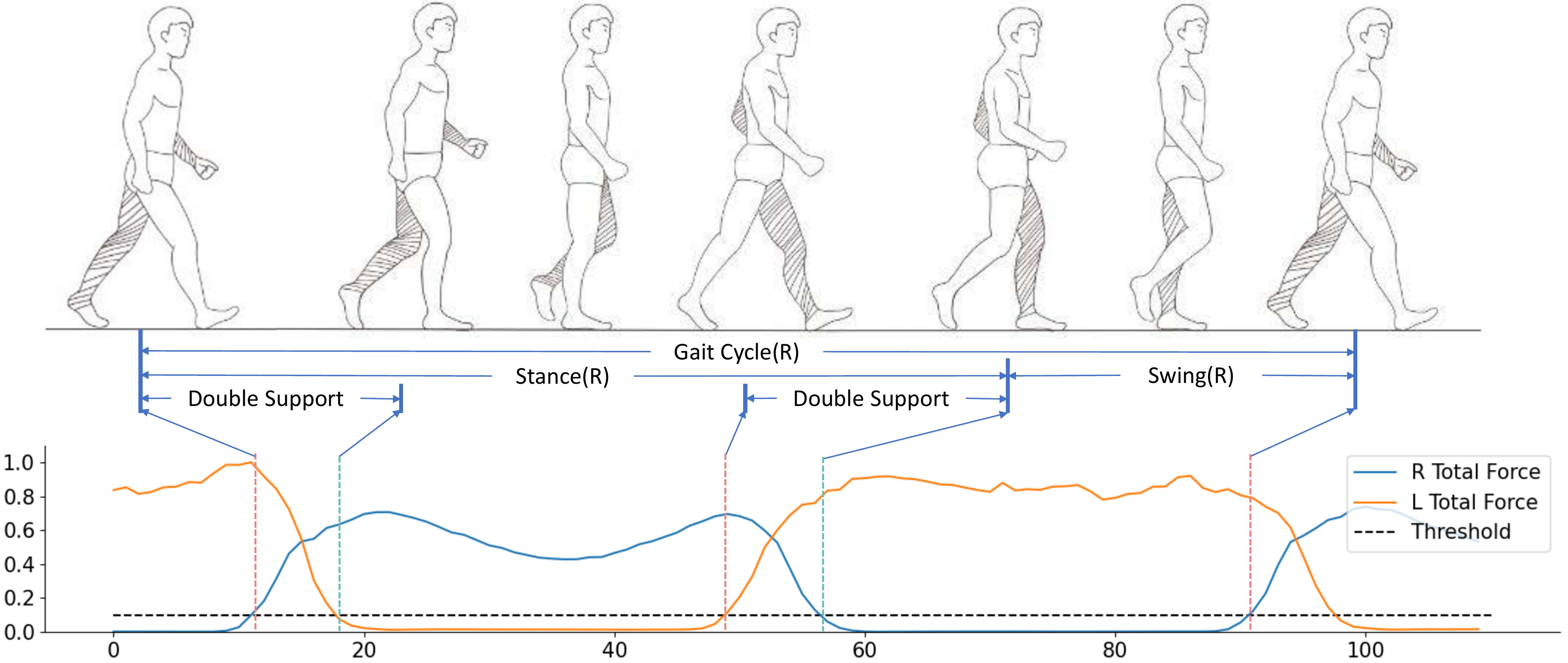}}
		\caption{The relation between the gait cycle and the total pressure}
		\label{fig:human_gait_cycle}
		\vspace{-0.3cm}
	\end{figure*}

Multiple frames are grouped into one step. As shown in Fig.~\ref{fig:human_gait_cycle}, one step is defined as the period from one heel strike to the next. The total force~($Fdes_1(t)$) is used to divide the whole walking test into many steps. The values are grouped into three clusters using K-means algorithms. Defining the centroids of the these clusters as: $x_1$, $x_2$, $x_3$~(in the ascending order), the threshold in Fig.~\ref{fig:human_gait_cycle} is calculated as:
	
	\begin{equation}
		threshold = 0.9 \times x_1 + 0.1 \times x_2
	\end{equation}

The grouped frame features are fed into the step features extraction layer and three types of step features are obtained: the phase duration features, the statistical features and the dual-modality fusion features.
	
	\subsubsection{Phase duration features}
	
	Each step is divided into the unipedal gait phase and the bipedal gait phase. The duration features for the $k^{th}$ step are:
	
	\begin{itemize}
		\item The duration of the whole cycle~($Spd_1(k)$), the stance phase~($Spd_2(k)$), the swing phase~($Spd_3(k)$) and the double support phase~($Spd_4(k)$). 
		\item Stance and swing phase ratios~($Spd_5(k)$, $Spd_6(k))$).
		\begin{equation}
			\begin{cases}
				Spd_5 = Spd_2 / Spd_1\\
				Spd_6 = Spd_3 / Spd_1
			\end{cases}
		\end{equation}
		\item The ratios of the double support phase to the previous and the next stance phase~($Spd_7(k)$, $Spd_8(k)$).
		\begin{equation}
			\begin{cases}
				Spd_7(k) = Spd_4(k) / Spd_2(k) \\
				Spd_8(k) = Spd_4(k) / Spd_2(k+1)
			\end{cases}
		\end{equation}
	\end{itemize}
	
	\subsubsection{Statistical features}
	taking the frame features within each step as the input, the statistical features of the $k^{th}$ step are calculated:
	
	\begin{itemize}
		\item Maximum, minimum, range, median, mean, standard deviation, coefficient of variation and information entropy~($Sst_1(k)$ to $Sst_8(k)$);
		\item The number of peaks~($Sst_9(k)$) and valleys~($Sst_{10}(k)$);
		\item Skewness~($Sst_{11}(k)$):
		\begin{equation}
			Sst_{11}(k) = \frac{\sum\limits_{t}^{T}(Fdes_i(t) - Sst_5(k))^3}{T \cdot Sst_6(k)^3}
		\end{equation}
		\item Kurtosis~($Sst_12(k)$):
		\begin{equation}
			Sst_{12}(k) = \frac{\sum\limits_{t}^{T}(Fdes_i(t) - Sst_5(k))^4}{T \cdot Sst_6(k)^4}
		\end{equation}
	\end{itemize}
	
	\subsubsection{Dual-modal fusion features}
	\label{subsubsec:fusionfeature}
	The hemiplegic patients are impaired in walking. Their performance under certain extreme conditions might be different from that of the healthy subjects, e.g., the acceleration of the hemiplegic side in the left-right direction when the total force on the non-hemiplegic side is maximum~(the foot strikes the ground or the toes go off the ground). In order to capture the subject's performance under such conditions, we extend this method: when one frame parameter reaches its maximum or minimum value, the value of another frame parameter is taken as one dual-modality fusion feature. Given two frame parameters $Fdes_{i}(t)$ and $Fdes_{j}(t)$, $i \neq j$, their dual-modality fusion features are defined as:
	
	\begin{equation}
		\begin{cases}
			Sdf_{1,i,j}(k) = Fdes_{j}(\alpha) \\ 
			\alpha = argmax \; Fdes_{i}(t), t\in(1, T)
		\end{cases}
	\end{equation}
	
	\begin{equation}
		\begin{cases}
			Sdf_{2,i,j}(k) = Fdes_{j}(\beta) \\ 
			\beta = argmin \; Fdes_{i}(t), t\in(1, T)
		\end{cases}
	\end{equation}

There are 40 frame features for left and right foot respectively, we thus get $C_{80}^{2}=3160$ dual-modality fusion features for each step.

	\subsection{Whole features extraction}
	\label{subsec:Whole_features}
	
	Taking the step features within one test, the whole features are their statistics and the bipedal symmetry coefficients.
	
	\subsubsection{Statistical features}
	the maximum, minimum, range, median, mean, standard deviation, coefficient of variation, information entropy, the number of peaks and valleys, skewness and kurtosis of the step features ~($Wst_1$-$Wst_{12}$).
	
	\subsubsection{Symmetry features}
	 hemiplegic patients have different motor abilities on the hemiplegic side and the other side. Define the whole statistical features of the left foot and the right foot as $Wst_i^L$, $Wst_i^R$~($i \in [1, 12]$), the symmetry coefficient~($Wsy_{i}$) is:
	\begin{equation}
		Wsy_{i} = 1 - \frac{min(Wst_i^L, Wst_i^R)}{max(Wst_i^L, Wst_i^R)}
	\end{equation}
	
	 In order to improve the effectiveness of whole features in classification and regression, the whole features are normalized using Z-score standardization.
	
	\begin{equation}
		z = (x - u)/ \sigma 
	\end{equation}
where $u$ and $\sigma$ are the mean and the standard variance of $x$.
	
	\subsection{Feature selection}
	\label{subsec:feature_selection}
	
	After the above workflow, 270114 features are obtained, that are much more than the test subjects' number. Feature selection is adopted to overcome the overfitting problem. 
	
	There are three types of feature selection methods: the filter, the wrapper, and the embedding method~\cite{guyon2003introduction}. We choose the embedding method because it combines the advantages of the other two methods: the short computation time of the filter method and the optimal feature set that can be generated by the wrapper method for a specific selector.
	
	The embedding method include several steps: 1) specify a particular selector~(e.g., decision tree, support vector machine), 2) training the corresponding task~(classification or regression) using the given feature set~(in our case, the whole features), 3) obtain each feature's importance to the result, 4) rank and select the $k$ most important features.
	
	 In order to obtain a stable ranking, the data are randomly divided into five portions, one portion is excluded each time, the rest four portions are used for training. To conduct a smaller selected feature set to mitigate the overfitting issues, each feature from the obtained feature set will be selected in a descending order based on the mean feature importance of the five portions and calculated the Pearson correlation coefficients with every feature included in the selected feature set. If all the correlation results are smaller than 0.9, the feature will be added to the selected set until the size of the selected feature set is up to 10. Finally, we get an efficient and well-performing feature set. 
	
	\section{Results}
	\label{sec:results}
	A two-step approach is designed.  First, the subject is classified as ''patient'' or ''healthy''. If ''patient'', the lower limbs' muscle strength is predicted.
	
	\subsection{Patient-healthy classification}
	\label{subsec:classify}
	
	The first step is to classify the subject. We study the different test combinations: \textit{Straight~(Str)}, \textit{Right turning~(RT)}, \textit{Left Turning~(LT)}, \textit{RT+LT}, and \textit{All~(Str+RT+LT)}. When multiple tests are included~(viz. \textit{RT+LT}, or \textit{All}), the probabilities in different tests are summed up. The class with the higher probability is taken as the classification  result.
	
	We treat every test of one person as a sample. The number of samples~($N_{sample}$) for a given test combination is then: 
	\begin{equation}
	    N_{sample} = (N_{p} + N_{h}) \times N_{test}
	\end{equation}
where $N_{p}=23, N_{h}=17$ is the number of the patients and the healthy subjects, and $N_{test}$ is the number of test(s) in the combination). The sample numbers of the five test combinations are then 40, 40, 40, 80 and 120. 
	
	We study the effects of using different test combinations, feature selectors~(C-DT: decision tree, C-SVM: support vector machine, C-RF: random forest and C-LR: logistic regression) and classifiers~(DT: decision tree, SVM: support vector machine, KNN: K-nearest neighbor, RF: random forest, LR: logistic regression and MLP: multi-layer perceptron). Five-fold cross-validation is adopted, with no subject crossover between the training and test sets, and the results are in Table~\ref{table:classification_feature_selections}. It can be see that using decision tree as the selector and support vector machine as the classifier, the classification accuracy reaches 100\% when the data from \textit{RT} tests are used.
	
	\begin{table}[!ht]
		\setlength\tabcolsep{1pt}
		\caption{Healthy-patient classification result~(F1-Score), with test(s) combination, feature selector and classifier evaluated. The bolded part is the optimal result}
		\centering
		\begin{tabular}{|c|c|c|c|c|c|}
			\hline
			\diagbox{Test(s)}{Selector} & \textbf{C-DT} & C-SVM & C-RF & C-LR  \\ \hline
			\textit{Str} & \textbf{0.979(RF)} & 0.846(KNN) & 0.936(RF) & 0.880(KNN)  \\ \hline
			\textit{RT} & \textbf{1.000(SVM)} & 0.875(RF) & 0.957(RF) & 0.917(RF)  \\ \hline
			\textit{LT} & \textbf{0.979(KNN)} & 0.870(LR) & 0.955(SVM) & 0.917(MLP)  \\ \hline
			\textit{RT+LT} & \textbf{0.979(RF)} & 0.958(RF) & 0.909(RF) & 0.936(RF)  \\ \hline
			\textit{All} & \textbf{0.979(KNN)} & 0.936(RF) & 0.936(RF) & 0.898(RF)  \\ \hline
		\end{tabular}
		\label{table:classification_feature_selections}
	\end{table}
	
	We then take the data from the \textit{RT} combination and study the effect of the sensing modalities, including: 1) IMU, as if there were only IMUs on the shoe; 2) Pressure, as if there were only the pressure insole; 3) IMU+Pressure, both sensing modalities are available, but the dual-modality fusion features are excluded, this is to test the effects of our newly proposed dual-modality fusion features; 4) All: all sensing modalities and features are included. Again multiple classifiers are tested, and the results are shown in Table~\ref{table:classification_dataset_combinations}. It can be seen that the classification of features calculated using only pressure is poor, and We can only use IMU to classify patients with healthy subjects.
	
	\begin{table}[!ht]
		\setlength\tabcolsep{2pt}
		\caption{Sensing modalities evaluation~(selector: C-DT)}
		\centering
		\begin{tabular}{|c|c|c|c|c|}
			\hline
			Dataset & The best classifier & Accuracy & Precision & F1-score  \\ \hline
			IMU & SVM & 0.975 & 1.000 & 0.978  \\ \hline
			Pressure & SVM & 1.000 & 1.000 & 1.000  \\ \hline
			IMU+Pressure & SVM & 1.000 & 1.000 & 1.000  \\ \hline
			All & SVM & 1.000 & 1.000 & 1.000  \\ \hline
		\end{tabular}
		\label{table:classification_dataset_combinations}
	\end{table}
	
	\subsection{Regression result for muscle strength estimation}
	\label{subsec:regression}
	
	After classifying the subjects, further regressions are performed to predict the patients' lower limbs' muscle strength. 
		
	We study the effects of using different test combinations~(\textit{Str}, \textit{RT}, \textit{LT}, \textit{RT+LT} and \textit{All}), feature selectors~(R-DT: decision tree, R-SVM: support vector machine, R-RF: random forest, R-SR: stochastic gradient descent regression) and regressors~(DT: decision tree, SVM: support vector machine, KNN: K-nearest neighbor, RF: random forest, GB: gradient boosting tree and MLP: multi-layer perceptron). When there are multiple tests in the combination, the average of the predicted muscle strength from different tests is taken. Leave-one-out method is adopted, with no subject crossover between the training and test sets. The difference to the ground truth is measured by Mean Absolute Error~(MAE), Root Mean Square Error~(RMSE) and Maximum Error~(ME):
	
	\begin{equation}
		\begin{cases}
			MAE = {\sum_{i=1}^n|y_i - x_i|}/ {n} \\ 
			RMSE = \sqrt{{\sum_{i=1}^n(y_i - x_i)^2}/ {n}} \\
			ME = max(|y_i - x_i|)
		\end{cases}
	\end{equation}
	
\begin{figure}[htbp]
		\centerline{\includegraphics[width=0.5\textwidth]{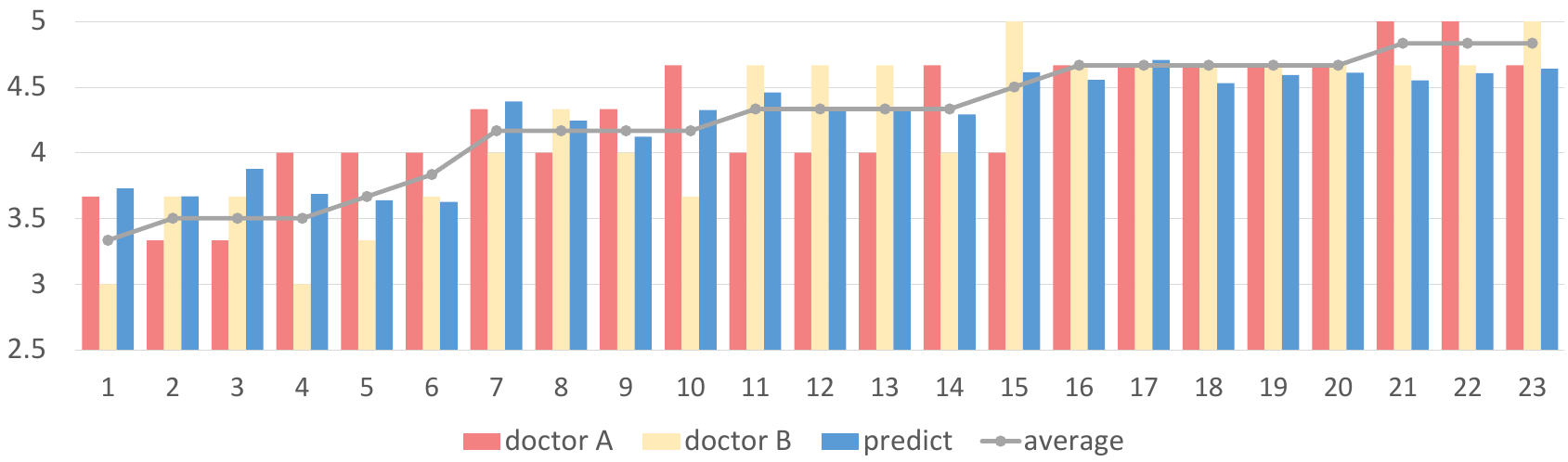}}
		\caption{The predicted results based on left-turn and right-turn experiments, compared with the evaluation values of two doctors.}
		\label{fig:doctor_muscle_strength}
	\end{figure}

The results are shown in Table~\ref{table:regression_feature_selections}.	It can be seen that using decision tree as the selector and random forest as the classifier, the best MAE is 0.138 when the tests \textit{RT+LT} are used. We then fix the selector and compare the MAE's and ME's of different test combinations. The results are given in Table.~\ref{table:regression_combinations_of_experiments}. It can be seen that although \textit{RT+LT} is of the smallest MAE, its ME is much higher than that of \textit{RT}, whose MAE is  the second optimum and just a little higher than \textit{RT+LT}. Because the estimated muscle strength should maintain not only a low overall prediction error, but also a low error for each patient, \textit{RT} is taken as the optimum. Why \textit{RT} is the best test may be explained by the fact that 78\%~(18/23) of the patients are right-sided hemiplegic and thus show a greater difference during right turning. We then compare the algorithm's performance with that of the two physicians. The results are shown in Fig.~\ref{fig:doctor_muscle_strength}. It can be seen that for this specific dataset, the algorithm performs even better in giving the muscle strength than the physicians as an individual~(algorithm: MAE 0.143, RMSE 0.178, ME 0.395; physicians: MAE 0.217, RMSE 0.269, ME 0.500). 
	
	\begin{table}[!ht]
		\setlength\tabcolsep{2pt}
		\caption{Muscle strength regression results~(MAE), with test(s) combination, feature selector and classifier evaluated. The bolded part is the optimal result}
		\centering
		\begin{tabular}{|c|c|c|c|c|c|}
			\hline
			\diagbox{Test(s)}{Selector}  & \textbf{R-DT} & R-SVM & R-RF & R-SR  \\ \hline
			\textit{Str} & \textbf{0.167(GB)} & 0.348(KNN) & 0.253(RF) & 0.353(GB)  \\ \hline
			\textit{RT} & \textbf{0.143(RF)} & 0.360(GB) & 0.201(RF) & 0.391(GB)  \\ \hline
			\textit{LT} & \textbf{0.214(RF)} & 0.211(GB) & 0.211(RF) & 0.324(RF)  \\ \hline
			\textit{RT+LT} & \textbf{0.138(RF)} & 0.300(GB) & 0.194(KNN) & 0.327(GB)  \\ \hline
			\textit{All} & \textbf{0.158(RF)} & 0.313(GB) & 0.172(SVM) & 0.360(DT)  \\ \hline
		\end{tabular}
		\label{table:regression_feature_selections}
	\end{table}

	\begin{table}[!ht]
		\caption{Finding the optimum test combination for the regression task~(R-DT as the selector). The bolded part is the minimum error}
		\centering
		\begin{tabular}{|c|c|c|c|c|}
			\hline
			Test(s) & Regressor & MAE &RMSE & ME \\ \hline
			\textit{Str} & GB & 0.167 &0.239 & 0.614 \\ \hline
			\textit{RT}& RF & \textbf{0.143} & \textbf{0.178} & \textbf{0.395} \\ \hline
			\textit{LT} & RF & 0.214 & 0.241 & 0.450 \\ \hline
			\textit{RT+LT} & DT & \textbf{0.138} & \textbf{0.174} & 0.441 \\ \hline
			\textit{All} & RF & 0.158 & 0.198 & 0.503 \\ \hline
		\end{tabular}
		\label{table:regression_combinations_of_experiments}
	\end{table}
	
	\subsection{Features importance analysis}
	\label{subsec:feature}
	
	To better understand which features play more important role in estimating the muscle strength, we used the forward search algorithm. The initial feature set is empty, and the best result is obtained by adding one non-repeating optimal feature at a time. The names and results of the 10 features added to the feature set in order are shown in the table \ref{table:the_names_and_results}. Three types of features are selected: 1) \textit{$Fdes_i$}\_Max(Min)Time\_\textit{$Fdes_j$}\_\textit{$S_1$}, 2) \textit{$Fdes_i$}\_Max(Min)Time\_\textit{$Fdes_j$}\_\textit{$S_1$}\_Asymmetry and 3) \textit{$Fdes_i$}\_\textit{$S_1$}\_\textit{$S_2$}. Here \textit{$Fdes_i$} and \textit{$Fdes_j$} denote a frame feature~(e.g., acceleration in the y-direction of the right forefoot~($Fdes_{30}(t)$) is named RForeGyroY). \textit{$S_1$} and \textit{$S_2$} denote a statistical calculation~(e.g., mean, maximum). The three types are then: 
	1) the \textit{$S_1$} operation on \textit{$Fdes_j$}'s values at the maximum~(minimum) time of value \textit{$Fdes_i$} within each step; 
	2) the asymmetry coefficient of \textit{$Fdes_i$}\_Max(Min)Time\_\textit{$Fdes_j$}\_\textit{$S_1$} between the left and right feet; 
	and 3) perform the \textit{$S_1$} operation on \textit{$Fdes_i$}'s values within a step, then perform the \textit{$S_2$} operation on all \textit{$Fdes_i$}\_\textit{$S_1$} in the test.
	
	\begin{table}[!ht]
	\setlength\tabcolsep{1pt}
	\caption{The names and results of the features added to the feature set in order}
    \centering
    \begin{tabular}{|c|c|c|c|c|}
    \hline
        No. & Feature names & MAE & RMSE  & ME  \\ \hline
        1 & LBackGyroZ\_MinTime\_LForeGyroY\_Range  & 0.242 & 0.293  & 0.692  \\ \hline
        2 & LImageMax\_MinTime\_RBackGyroY\_SD & 0.192 & 0.267  & 0.778  \\ \hline
        3 & LBackGyroXyz\_MaxTime\_LBackAccXyz\_SD & 0.164 & 0.220  & 0.582  \\ \hline
        4 & LForeGyroXyz\_MaxTime\_LBackAccX\_ & 0.133 & 0.169  & 0.410  \\ 
        ~ & Entropy\_Symmetry & ~ & ~ & \\ \hline
        5 & RMatCentreSpeed\_MinTime\_RForeAccZ\_ & 0.128 & 0.158  & 0.333  \\
        ~ & Median & ~ & ~ & \\ \hline
        6 & RImageCV\_MaxTime\_RForeGyroX\_Max & 0.126 & 0.149  & 0.310  \\ \hline
        7 & RForeGyroY\_MaxTime\_RForeAccX\_Median & 0.122 & 0.163  & 0.34  \\ \hline
        8 & RImageMean\_MaxTime\_LBackAccX\_SD & 0.148 & 0.193  & 0.458  \\ \hline
        9 & RImageHu5\_Skewness\_Min & 0.142 & 0.179  & 0.365  \\ \hline
        10 & LImageSD\_Skewness\_Min & 0.168 & 0.200  & 0.420  \\ \hline
    \end{tabular}
    \label{table:the_names_and_results}
    \end{table}
    
    \begin{figure}[htbp]
		\centerline{\includegraphics[width=0.5\textwidth]{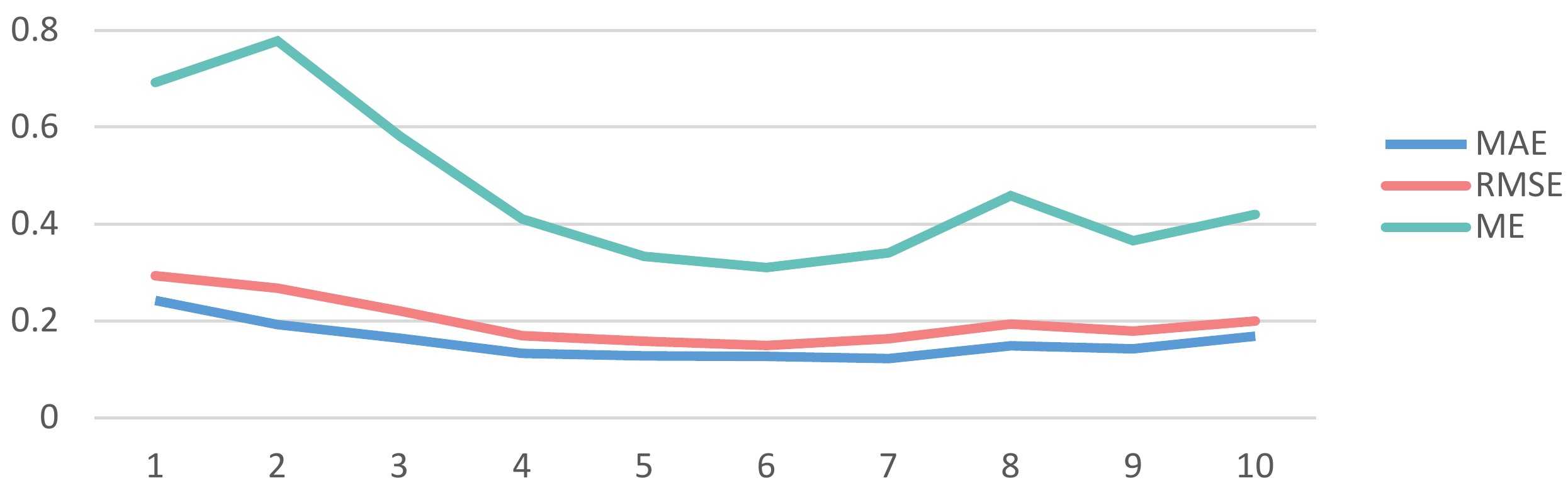}}
		\caption{The experimental results using forward search algorithm}
		\label{fig:feature_importance}
		\vspace{-0.2cm}
	\end{figure}
	
It can be seen that eight of ten top features are the newly proposed dual-modality fusion features. This demonstrates that the great importance of our new features to the muscle strength estimation. When using 4-7 features, the results have leveled off, while at greater than 7 features, the error starts to increase and may start to overfit. This suggests that using 4-7 features is most appropriate in this experiment.
	
	\section{Conclusion and Discussion}
	\label{sec:conclusion}
	
	In this paper, we demonstrate the possibility of using a pair of smart shoes to non-intrusively and objectively assess the hemiplegic patients' lower limbs' muscle strength. In doing so, we designed the extended 5m walk test protocol, also include the right and left turnings, which are proven to be useful in the assessment. We create a feature set to describe the characteristics of the walking, including the newly proposed dual-modality fusion features, which are also proven to be useful. Based on the data gathered from 23 patients and 17 healthy subjects, a 100\% classification result of ''healthy-patient'' is achieved. For estimation the muscle strength, regression methods are evaluated, the algorithm's best performance is MAE 0.143, RMSE 0.178 and ME 0.395, both better than that of the physicians as an individual~(MAE 0.217, RMSE 0.269 and ME 0.500). 
		
	There are still rooms for improvement, for example, finding out on which side is the hemiplegic leg, or segmenting the test data into going straight and turning. These operations can create new features that might be useful. Estimation of the other subjective medical assessment values~(e.g., NHISS scores) can also be carried out using the same dataset. Through long-term data acquisition at home, the rehabilitation results could be evaluated. We believe that smart shoes could become a useful tool in quantitative assessment of the lower limbs' muscle strength for hemiplegic patients.

 	\section{Acknowledgment}
 	\label{sec:acknowledgment}
 This work is approved by the Bio-medical Ethics Committee of University of Science and Technology of China~(reference number: 2021KY-054) and supported by ”the Fundamental Research Funds for the Central Universities” (Grant No. 2150110020).
	
    \bibliographystyle{gbt7714-numerical}
	\normalem
	\bibliography{reference.bib}
	
\end{document}